\documentclass[aps,prl,reprint,amsmath,amssymb,graphicx,longbibliography,superscriptaddress]{revtex4-2}

\usepackage{bm}
\usepackage{amsmath}
\usepackage[normalem]{ulem}
\usepackage[colorlinks=true,allcolors=blue,bookmarks=false,pdfusetitle]{hyperref}
\usepackage{url}
\usepackage{xcolor} 
\usepackage{graphicx}
\usepackage{breqn}
\usepackage{braket}
\usepackage{cases}
\usepackage[utf8]{inputenc}
\usepackage[T1]{fontenc}
\usepackage[overload]{empheq}
\usepackage{cleveref} 

\usepackage{times}

\usepackage{tikz}

\makeatletter
\let\cat@comma@active\@empty
\makeatother

\usepackage{booktabs}
\providecommand{\numberTblEq}[1]{\refstepcounter{tblEqCounter}\label{#1}\thetag{\thetblEqCounter}}

\begin{document}
%\nocite{*}

\title{Quantum simulation of thermodynamics: Maxwell relations for pair correlations }

\author{F. Rist}
\author{R. S. Watson}
\affiliation{School of Mathematics and Physics, University of Queensland, Brisbane,  Queensland 4072, Australia}
\author{H. L. Nourse}
\affiliation{School of Chemistry, University of Sydney, Sydney, NSW 2006, Australia}
\author{B. J. Powell}
\email{powell@physics.uq.edu.au}
\author{K. V. Kheruntsyan}
\email{karen.kheruntsyan@uq.edu.au}
\affiliation{School of Mathematics and Physics, University of Queensland, Brisbane,  Queensland 4072, Australia}

\date{\today{}}

\begin{abstract}
\noindent 

Quantum simulators hold enormous promise for advancing the modelling of materials and understanding emergent physics, such as high temperature superconductivity and topological order. While correlation functions are, typically, straightforward to measure in quantum simulators, thermodynamic properties are not. This limits our ability to directly compare the results of quantum simulations to experiments on the materials being modelled. Maxwell relations are an extremely powerful tool for characterising complex materials, as they enable the determination of challenging-to-measure thermodynamic properties from more accessible ones. Here, we introduce generalised Maxwell relations that relate every thermodynamic quantity to a single local correlation function. We illustrate their utility by deducing the thermodynamic properties of several iconic quantum many-body models from pair correlation functions using the generalised Maxwell relations. We show that this {universal} approach is readily accessible in quantum simulators and suggest applications to condensed matter systems where thermodynamic measurements are challenging, such as atomically thin materials. 
\end{abstract}

\maketitle

Condensed matter physics is entering an era of unprecedented control over the design and synthesis of quantum materials. Engineered materials such as metal-organic frameworks \cite{Yaghi}, twistronics devices \cite{Cao2018,Xia2025} and atomically thin 2D materials \cite{Kalmutzki2018,Lowe,Takenaka2021} have emergent properties that may enable breakthrough applications from tackling climate change to the quantum technology revolution. However, emerging properties are difficult to predict \cite{Powell}, making it challenging to exploit this structural control to achieve new functionalities.

\begin{figure*}[tbp] 
    \includegraphics[width=17.8cm]{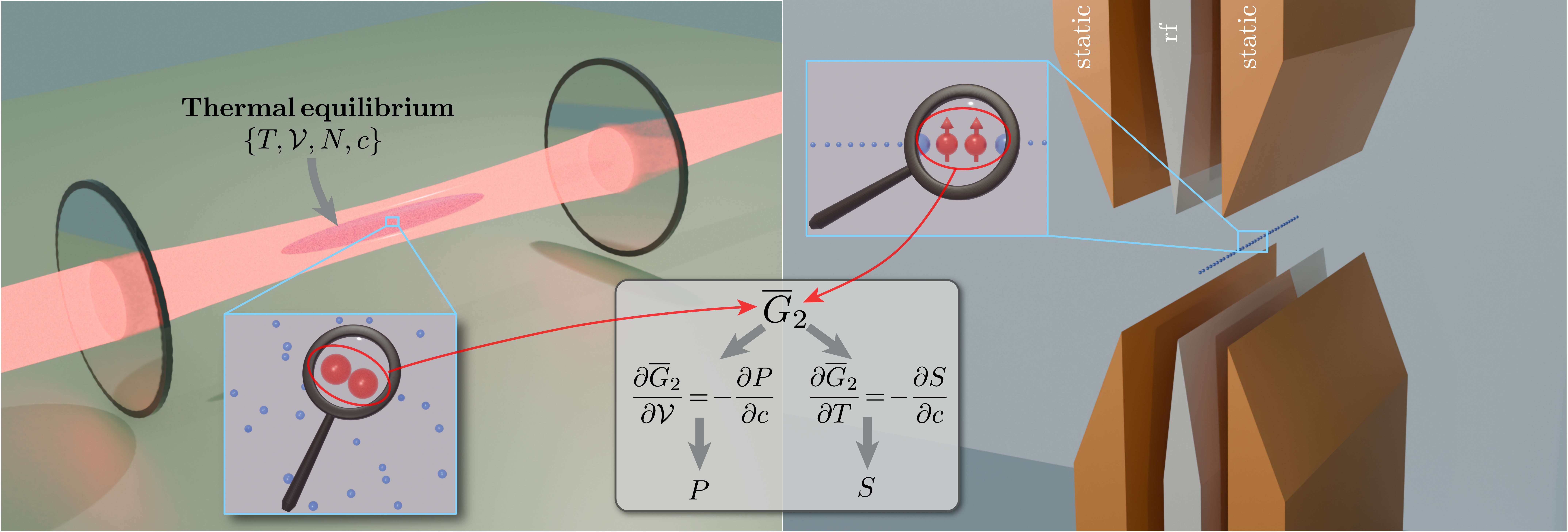}
    \caption{An illustration of the key idea behind this work: a quantum gas microscope (left) or an ion trap (right) measures the two-atom or two-spin integrated correlation $\overline{G_2}$ as a function of the interaction strength $c$ and other thermodynamic parameters such as temperature ($T$), system size ($\mathcal{V}$), number of particles ($N$), etc.; the Maxwell relations derived here allow for an arbitrary thermodynamic quantity---such as pressure $P$, entropy $S$, chemical potential, heat capacity, magnetization, etc.---to be deduced from the measured  $\overline{G_2}$. 
    \label{fig:Illustration}}
\end{figure*}

One possible approach to designing new quantum materials is to harness the rapidly increasing potential of quantum simulators---precursors to general purpose quantum computers that allow the accurate simulation of a single model \cite{Fauseweh2024}. The most advanced quantum simulators include ultracold atoms in optical lattice potentials \cite{Gross_Bloch_2017,Greiner_microscope_correlations_2016,Bloch_microscope_correlations_2025} (Fig.~\ref{fig:Illustration}a) or tweezer arrays \cite{Ebadi2021,Semeghini2021}, quantum dot arrays \cite{Hensgens2017}, superconducting circuits \cite{Kjaergaard,Salathe} and ion traps \cite{Munro2021,Guo2024,Kim2010,Lanyon} (Fig.~\ref{fig:Illustration}b). All offer precise control of the underlying Hamiltonian, rarely possible in traditional condensed matter systems. Therefore, they are ideal test beds for theory and provide platforms for understanding and engineering emergent physics. Superfluidity, topological phases, and metal-insulator transitions have been demonstrated in these devices. However, in quantum simulators, it is much easier to measure correlation functions \cite{Gross_Bloch_2017,Greiner_microscope_correlations_2016,Bloch_microscope_correlations_2025,Ebadi2021,Semeghini2021,Hensgens2017,Munro2021,Guo2024,Kim2010,Lanyon,Salathe,Kjaergaard}  than the thermodynamic variables that have provided invaluable insights in traditional condensed matter systems, from uncovering new states of matter to understanding universality in phase transitions. Furthermore, as thermodynamic measurements remain a mainstay of materials characterisation and condensed matter physics, this limits our ability to compare quantum simulations to experiments on the materials being modelled.

Here, we propose a new \emph{universal} method of extracting \emph{any} thermodynamic quantity from measurements of a single correlation function. To do this we generalise Maxwell relations---which allow one thermodynamic quantity to be determined from measurements of another \cite{Callen_book}---to include correlation functions;  Fig.~\ref{fig:Illustration}. 
This contrasts with previous attempts to extract thermodynamics from correlation functions \cite{Werner2005,Watson2024}  or Tan's contact  \cite{chen2014critical,Liu2020} that are formulated for specific models or require (experimentally challenging) measurements of high-energy tails of momentum distributions. Our method requires only measurements of local pair correlation, making it broadly applicable to a wide range of systems. Thus, this work opens the door to widespread measurements of the thermodynamic properties in quantum simulators.

We validate the generalised Maxwell relations, by using them to calculate the properties of well-studied problems including Bose and Fermi gases, Mott transitions, and magnetic quantum phase transitions. Extracting thermodynamic properties from experimentally measured correlation functions would follow the same logic. Thus, this demonstrates the utility of our approach as a tool for both theory and experiment.

We also use the generalised Maxwell relations to calculate an analytical expression for the entropy of a strongly interaction Fermi gas, demonstrating the ease with which new theory can be developed using this method. 

Finally, we find that the integrated correlation function, introduced below and key to our approach, is a thermodynamic variable. We demonstrate that this can therefore be used to probe phase transitions in materials where it is difficult or impossible to measure conventional thermodynamic properties, such as 2D materials.

~

\noindent\textbf{Model systems and correlation functions}\\
\noindent The derivation of the generalised Maxwell relations is, almost embarrassingly, simple but requires a few definitions, which we give in this section.

We begin by considering a quantum many-body interacting system with pairwise interactions, characterised by the Hamiltonian
\begin{equation}
	\hat{H}=\hat{H}_0+c\,\hat{\overline{G_2}}.
	\label{eq:Hamiltonian-general}
\end{equation}
Here, $\hat{H}_0$ is the non-interacting Hamiltonian, $c$ parametrises the strength of two-body interactions, and $\hat{\overline{G_2}}$ is a two-body operator, which we will refer to as an integrated pair correlation for reasons that will become clear shortly.

For many paradigmatic many-body models the operator $\hat{\overline{G_2}}$ can be expressed as an integral or sum over creation, annihilation, or spin operators, such as:
\begin{subequations}
	\label{eq:def}
	\begin{align}[left ={ \hat{\overline{G_2}} \!=\! \empheqlbrace}]
	& \!\iint \!  d\mathbf{r} d\mathbf{r}' \hat{\Psi}^{\dagger}(\mathbf{r}) \hat{\Psi}^{\dagger}(\mathbf{r}') f(\mathbf{r},\mathbf{r}') \hat{\Psi}(\mathbf{r}') \hat{\Psi}(\mathbf{r}).\label{eq:bosons}\\
		& \sum\limits_{j} \hat{c}^{\dagger}_{j,\uparrow} \hat{c}^{\,}_{j,\uparrow} \hat{c}^{\dagger}_{j,\downarrow} \hat{c}^{\,}_{j,\downarrow},\label{eq:FermiHubbard}\\
		& \sum_{j}\hat{S}_j^z\hat{S}_{j+1}^z.\label{eq:Ising} 
	\end{align}
\end{subequations}
Equation~\eqref{eq:bosons} describes the interactions in a gas of identical bosons, annihilated (created) at position
 $\mathbf{r}$  by the field operator $\hat{\Psi}^{(\dagger)}(\mathbf{r})$, with $f(\mathbf{r},\mathbf{r}')$ describing the spatial dependence of the interaction. 
Equation~\eqref{eq:FermiHubbard} is the Fermi-Hubbard interaction between fermions annihilated (created) at site $j$ with spin $\sigma$ by the operator $\hat{c}_{j,\sigma}^{(\dagger)}$.  
Equation~\eqref{eq:Ising} is the Ising interaction between the {$z$-components, $\hat{S}^z_j$, of adjacent spins on a lattice}.

The expectation value 
\begin{equation}
	\overline{G_2} \equiv \langle \hat{\overline{G_2}}\rangle,
	\label{eq:G2bar}
\end{equation} 
is proportional to an integral or a sum over local correlation functions. For example, the Lieb-Liniger model  \cite{Lieb-Liniger-I} {(see Methods)} of a 1D Bose gas assumes contact interactions, $f(x,x')\!=\!\delta(x-x')$. For a uniform system of length $\mathcal{V}$  the integrated correlation is proportional to the local two-particle correlation function: $\overline{G_2}\!=\!\mathcal{V}G^{(2)}\!=\!\mathcal{V}\langle \hat{\Psi}^{\dagger}(x) \hat{\Psi}^{\dagger}(x) \hat{\Psi}(x) \hat{\Psi}(x)\rangle$. 

~

\noindent\textbf{Maxwell relations for pair correlation functions}\\
\noindent 
Standard Maxwell relations follow \cite{Callen_book} from the commutativity of mixed second derivatives of the Helmholtz free energy, $F$. Here we generalise this derivation by taking derivatives with respect to the interaction strength $c$ and a standard thermodynamic parameter,  $X$:
\begin{equation}
\left[\frac{\partial }{\partial c}\left(\frac{\partial F}{\partial X} \right)_{c,\dots}\right]_{X,\dots} = \left[\frac{\partial }{\partial X}\left(\frac{\partial F}{\partial c} \right)_{X,\dots}\right]_{c,\dots},
\label{eq:mixed}
\end{equation}
where the ellipsis represents all the other external parameters held fixed.
In the canonical formulation of statistical mechanics, which we adopt here (see Supplementary Information for formulation in other ensembles), $X$ can be, for example, the temperature, $T$, the total number of particles, $N$, or the system size, $\mathcal{V}$, (i.e., volume in 3D; area in 2D; or length in 1D). $Y= (\partial F/\partial X)_{c,\dots}$, is then a thermodynamic quantity, such as pressure, entropy, chemical potential, depending on the choice of $X$. Similarly,  $(\partial F/\partial c)_{X,\dots}=\overline{G_2}$ (see Methods and Supplementary Information). This yields the generalised Maxwell relation
\begin{equation}
\left(\frac{\partial Y}{\partial c}\right)_{X,\dots}= \left(\frac{\partial \overline{G_2}}{\partial X}\right)_{c,\dots},
\label{Maxwell-generic}
\end{equation}

Integrating the generalised Maxwell relation, equation \eqref{Maxwell-generic}, from some $c_0$ to $c$, where $Y(c_0)$ is known, allows one to evaluate $Y(c)$ from the measured or calculated derivative of the integrated correlation function:
\begin{equation}
    Y(c)=Y(c_0)
+\int_{c_0}^{c} \left(\frac{\partial \overline{G_2}(c')}{\partial X}\right)_{c',\dots}dc'.
\label{Maxwell-integral}
\end{equation}

The generalised Maxwell relations can straightforwardly be extended to treat second derivatives of the free energy. One finds (see Methods) that if $Y= \left( \partial^2 F/ \partial X\partial X'\right)_{c}$, {then}
\begin{align}
	\left( \frac{\partial Y}{\partial c}\right)_{X,X',\dots} = \left(\frac{\partial^2 \overline{G_2}}{\partial X\partial X'} \right)_{c,\dots}. 	\label{eq:2Maxwell-generic}
\end{align}
On integrating, one finds that
\begin{align}
	Y(c) = Y(c_0) + \int_{c_0}^{c} \left(\frac{\partial^2 \overline{G_2}(c')}{\partial X\partial X'} \right)_{c',\dots} dc'.
	\label{eq:2Maxwell-integral}
\end{align}
Second derivatives of the free energy, such as heat capacity, magnetic susceptibility, or isothermal compressibility play a crucial role in thermodynamics, for example, in probing continuous phase transitions (also known as second order phase transitions).

The differential (equations~\eqref{Maxwell-generic} and \eqref{eq:2Maxwell-generic}) and integral (equations~\eqref{Maxwell-integral} and \eqref{eq:2Maxwell-integral}) forms of the generalised Maxwell relations are the key results of this work. They imply that any thermodynamic quantity, $Y$,  can be determined from the measurements of the correlation function $\overline{G_2}$. 
To illustrate the utility of generalised Maxwell relations we discuss several possible choices for the thermodynamic quantities $X$ and $Y$ below; {the differential forms of the respective Maxwell relations are summarised in Table \ref{tab:Maxwell_all} (for the integral forms, see Supplementary Information).}

\newcounter{tblEqCounter} %create a counter
\setcounter{tblEqCounter}{\theequation} %at the start of the table, set the counter to equation numbering
\begin{table}[h!]
\caption{Maxwell relations between various thermodynamic quantities and the pair correlation $\overline{G_2}$.}
\label{tab:Maxwell_all}
\centering
%\scriptsize
\begin{tabular}{llc}
\toprule
Thermodynamic \\quantity & Maxwell relation with $\overline{G_2}$ & Eq.\\
\midrule
Pressure, $P$ & ${\displaystyle {\displaystyle \left(\frac{\partial P}{\partial c}\right)_{\mathcal{V},T,N}=-\left(\frac{\partial\overline{G_{2}}}{\partial\mathcal{V}}\right)_{c,T,N}}}$ & \numberTblEq{eq:pressure_table}\\ %set the equation number 
Entropy, $S$ & ${\displaystyle {\displaystyle \left(\frac{\partial S}{\partial c}\right)_{T,\mathcal{V},N}=-\left(\frac{\partial\overline{G_{2}}}{\partial T}\right)_{c,\mathcal{V},N}}}$ & \numberTblEq{eq:entropy_table}\\
Chemical potential, $\mu$\,\,\,\,\,\,\,& ${\displaystyle {\displaystyle \left(\frac{\partial\mu}{\partial c}\right)_{T,\mathcal{V},N}=\left(\frac{\partial\overline{G_{2}}}{\partial N}\right)_{c,T,\mathcal{V}}}}$ & \numberTblEq{eq:chemical_potential_table}\\ 
Magnetization, $\bm m$ & ${\displaystyle {\displaystyle \left(\frac{\partial\bm{m}}{\partial c}\right)_{\bm{h},T,\mathcal{V},N}=-\left(\bm{\nabla_{\bm{h}}}\overline{G_{2}}\right)_{c,T,\mathcal{V},N}}}$ & \numberTblEq{eq:magnetization_table}\\ 
Heat capacity, $C_{\mathcal{V}}$ & ${\displaystyle {\displaystyle \left(\frac{\partial C_{\mathcal{V}}}{\partial c}\right)_{T,\mathcal{V},N}=-T\left(\frac{\partial^{2}\overline{G_{2}}}{\partial T^{2}}\right)_{c,\mathcal{V},N}}}$ & \numberTblEq{eq:heat_capacity_table}\\ 
$\begin{array}{c}
\vspace{-0.2cm}\\
\hspace{-1.2cm}\text{Isothermal}\\
\hspace{-0.0cm}\text{compressibility, $\kappa_{T}$}\\
\vspace{-0.3cm}
\end{array}$
 & ${\displaystyle {\displaystyle \left(\frac{\partial\kappa_{T}^{-1}}{\partial c}\right)_{T,\mathcal{V},N}=\mathcal{V}\left(\frac{\partial^{2}\overline{G_{2}}}{\partial\mathcal{V}^{2}}\right)_{T,N,c}}}$ & \numberTblEq{eq:compressibility_table}\\ \bottomrule
\end{tabular}
\end{table}
\setcounter{equation}{\thetblEqCounter} %at the end of the table, set the equation numbering to the counter

~

\noindent\textbf{Bose and Fermi gases}\\
\textbf{\emph{Pressure}} ($Y \!\equiv \!-P$, $X\!\equiv\! \mathcal{V}$). For pressure,  $P\!=\!-(\partial F/\partial \mathcal{V})_{T,N,c}$,  {the differential form of the generalised  Maxwell relation \eqref{Maxwell-generic} takes the form of Eq.~\eqref{eq:pressure_table}, Table \ref{tab:Maxwell_all} (for the integral form, see Supplementary Information).}

The Lieb-Liniger model provides a simple context for exploring these relations. In the weakly interacting regime ($c \! \ll \! \hbar^2n/m$, where $m$ is the mass of the bosons and $n=N/\mathcal{V}$ is the  particle number density) and at zero temperature, one can use the mean-field approximation. In this regime, density-density fluctuations are uncorrelated, and therefore $\overline{G^{(2)}}\!=\!N^2/\mathcal{V}$.  The standard derivation of the non-trivial result $P(c)=-(\partial U_0/\partial \mathcal{V})_{T,N,c}=cn^2$ \cite{Pethick_Smith_book,kerr2024analytic}, at zero temperature, requires the calculation of the ground state energy, $U_0$. However, this result is much more easily obtained from the generalised Maxwell relation. Trivially, $\left({\partial \overline{G_2}}/{\partial \mathcal{V}}\right)_{c,T,N}=-N^2/\mathcal{V}^2=-n^2$. On choosing $c_0=0$ and noting that $P(0)=0$, {Eqs. \eqref{Maxwell-integral} and \eqref{eq:pressure_table}   immediately yield} $P(c)=cn^2$. (This derivation can be easily extended to 2D and 3D systems.) 
Thus, it is much easier to calculate the pressure from the correlation function than via standard thermodynamic approaches for this system.

\begin{figure*}[tbp] 
    \includegraphics[width=17.7cm]{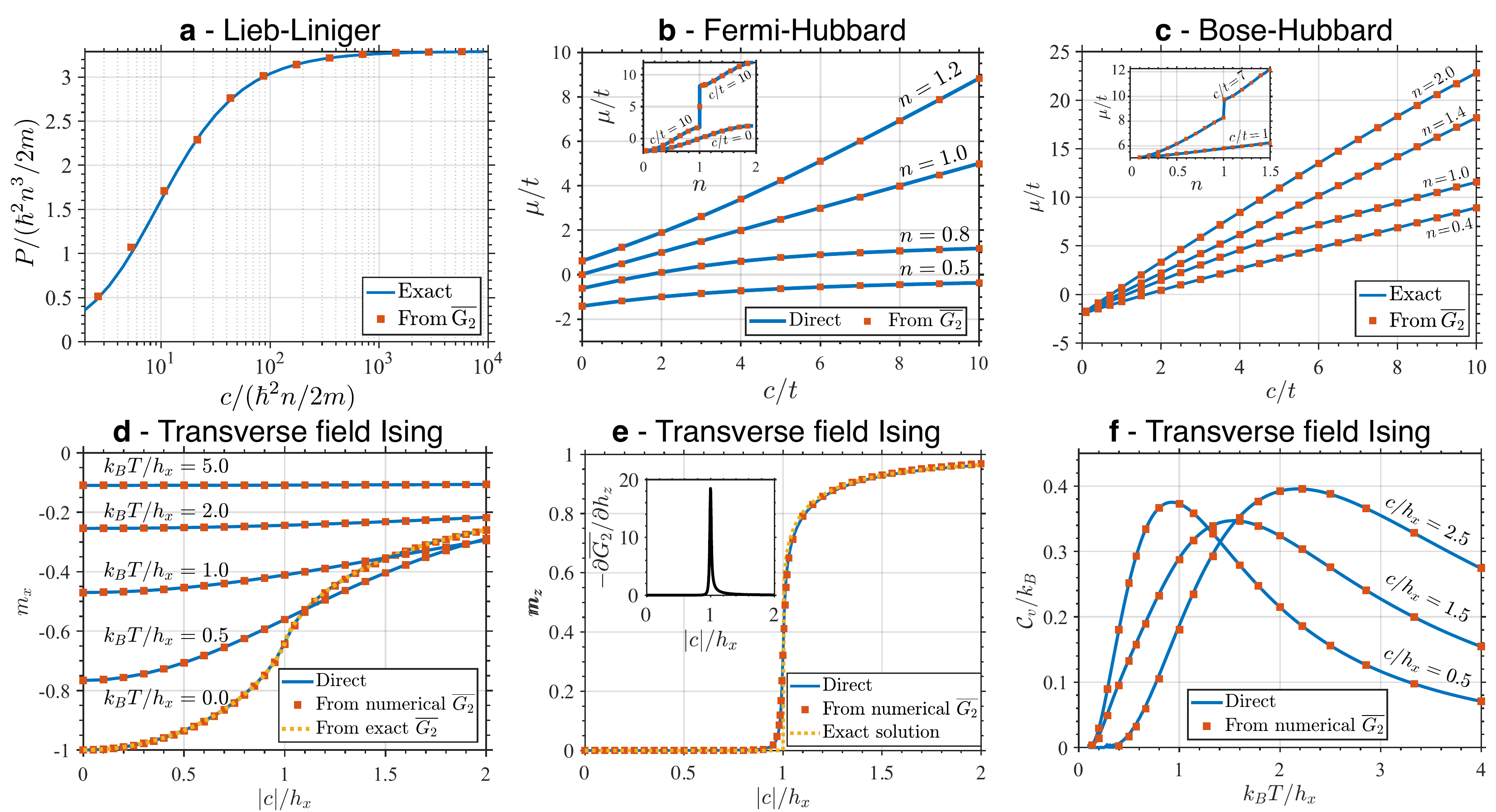}
    \caption{\label{fig:results}Comparison of thermodynamic quantities calculated from: (\emph{i}) the integrated correlation function, $\overline{G_2}$, using the Maxwell relations derived here and numerically calculated correlation functions (red squares), and (\emph{ii}) traditional methods of statistical mechanics either directly from our numerics (blue curves) or from exact results (yellow dashed curves). In all cases the agreement between the two methods is excellent. \textbf{a}, Pressure, $P$, of the strongly interacting Lieb-Liniger gas  at $T\!=\!0$ K. 
    	Chemical potentials, $\mu$, of the  1D, \textbf{b}, Fermi- and, \textbf{c}, Bose-Hubbard models, where  $c$ is the on-site  interaction and $t$ is the hopping integral. 
    	The insets show the chemical potential as a function of filling, $n$. The discontinuities in the chemical potentials when the number of particles equals the number of sites  ($n\!=\!1$) for large $c/t$, indicate (Mott) metal-insulator transitions \cite{Lieb_1968,LIEB_fermi_hubbard_review}. 
  	  	\textbf{d}, Transverse magnetisation, $m_x$ (see Supplementary Information for derivation of the exact result),
  		\textbf{e}, longitudinal magnetisation, $m_z$,  and \textbf{f}, heat capacity, $C_{\mathcal{V}}$, of the transverse field Ising model for field strength $h_x$. This demonstrates that, although $C_{\mathcal{V}}$ is the second derivative of the free energy, it can also be calculated essentially exactly from $\overline{G_2}$. Inset to \textbf{e}, the derivative of the pair correlation function $\overline{G_2}$ diverges at the critical point $|c|\!=\!h_x$, providing a signature of the continuous phase transition. This highlights that $\overline{G_2}$ is a thermodynamic variable.}
\end{figure*}

The generalised Maxwell equation approach also allows more straightforward calculations of thermodynamic properties in highly non-trivial regimes, for example the strongly interacting 1D Bose gas ($c>\hbar^2n/m$). In this limit thermodynamic properties can be calculated exactly using the thermodynamic Bethe ansatz (TBA) \cite{Yang-Yang}. 
We benchmark our Maxwell relations, Eqs. \eqref{Maxwell-integral} and \eqref{eq:pressure_table}, against the exact results by numerically evaluating $\overline{G_2}(c)$ via the density matrix renormalisation group (DMRG; see Methods and Supplementary Information) \cite{white1992density,schollwock2011density}. The two methods are in excellent agreement for the pressure (Fig.~\ref{fig:results}{a}) and the isothermal compressibility (Supplementary Information).

The equivalent model for {spin-$1/2$} fermions in 1D is known as the Yang-Gaudin model \cite{Yang1967,gaudin1967systeme} {(see Methods)}. Although this model is also exactly solvable using the TBA, approximate analytical results for thermodynamic quantities are scarce outside of the low-temperature Luttinger liquid regime~\cite{Zhao2009}. In the Supplementary Information we show that in the high-temperature limit, 
\begin{equation}\label{eq:YangGaudin_S}
	S \simeq S_{\mathrm{IFG}} -  \frac{N (1-\mathcal{P}^2)}{16\hbar } \sqrt\frac{\pi m}{2k_B T^3} nc^2.
\end{equation}
where $S_{\mathrm{IFG}}$ is the entropy of the ideal Fermi gas and $\mathcal{P}$ is the polarisation. The simplicity of the derivation of this result highlights the power of the generalised Maxwell equations for making progress with otherwise intractable problems.

~

\noindent\textbf{The Mott transition}\\
Mott insulators are of fundamental interest because of their role in high-temperature and unconventional superconductors \cite{LeeNW}, and because of potential  applications, including  ultrafast transistors,
volatile and non-volatile memories, and artificial neurons for neuromorphic computing \cite{Milloch,Lowe}. As such,  Mott insulators have been a  target for many quantum simulators \cite{
Greiner_microscope_correlations_2016,Bloch_microscope_correlations_2025}.

\textbf{\emph{Chemical potential}} ($Y \equiv \mu$, $X\equiv N$). The generalised Maxwell equation for the chemical potential, $\mu$, {takes the form of Eq.~\eqref{eq:chemical_potential_table} in Table \ref{tab:Maxwell_all}.} As a proof-of-principle study of the Mott transition, we calculated the zero temperature, local correlation functions, and the ground state energies of the 1D Fermi- and Bose-Hubbard models using DMRG (see Methods).
The chemical potentials of both models calculated from the generalised Maxwell relations are in excellent agreement with direct calculations using $\mu = \left( \partial F/\partial N\right)_{T,\mathcal{V},c}$, Figs.~\ref{fig:results}{b}, {c}.
For sufficiently strong interactions, a Mott gap opens at $n=1$ in both models (insets to Figs.~\ref{fig:results}{b}, {c}), indicated by a discontinuous increase in the chemical potential \cite{Ouyang_2013,Lieb_1968,LIEB_fermi_hubbard_review}. Importantly, the size of the Mott gap (i.e., the magnitude of the discontinuous increase in $\mu$) is also in excellent agreement for both methods, indicating that thermodynamics extracted from correlation functions reliably describe Mott physics in the strongly correlated regime.

~

\noindent\textbf{Magnetic insulators: the transverse field Ising model}\\
Capturing the physics near a phase transition requires extremely high accuracy experiments or calculations. Therefore, it is important to benchmark the accuracy of the thermodynamics extracted from correlation functions close to phase transitions as this provides a stringent test of the generalised Maxwell relations.
Therefore, we consider the transverse field Ising model, which is the simplest model that shows a quantum phase transition (i.e., a continuous phase transition at zero temperature) \cite{Sachdev_2011}, and has been a key target for quantum simulators \cite{Lanyon,Fauseweh2024,Kim2010,Munro2021,Guo2024,Salathe,Kjaergaard}.

\textbf{\emph{Magnetization}} ($Y=-\bm{m}$, $X=\bm{h}$). In an external magnetic field, $\bm{h}=\hat{\bf{i}} h_x+ \hat{\bf{j}} h_y+ \hat{\bf{k}} h_z$, the magnetization is $\bm{m}=\hat{\bf{i}} m_x+ \hat{\bf{j}} m_y+ \hat{\bf{k}} m_z=  -\mathcal{V}^{-1}(\bm{\nabla_{\bm{h}}}F)_{T,\mathcal{V},N, c}$, where $\bm{\nabla_{\bm{h}}}=\hat{\bf{i}} (\partial/\partial h_x)+ \hat{\bf{j}} (\partial/\partial h_y)+ \hat{\bf{k}} (\partial/\partial h_z)$. The relevant generalised Maxwell {relation is given in Eq.~\eqref{eq:magnetization_table}.}

We calculated the magnetization of the ferromagetic ($c\!<\!0$) transverse field Ising model (see Methods) by:
(i) {applying Eqs.~\eqref{Maxwell-integral} and \eqref{eq:magnetization_table}} to the $\overline{G_2}$ computed with thermal DMRG, where we take $c_0=0$, the limit of non-interacting spins;
(ii) applying Eqs.~\eqref{Maxwell-integral} and \eqref{eq:magnetization_table}  to the $\overline{G_2}$ calculated exactly at $T=0$ (Supplementary Information); and
(iii) directly evaluating  $m_x = (1/\mathcal{V}) \sum_{j=1}^{\mathcal{V}} \langle \hat{S}_j^x\rangle$ using  DMRG. 
All three methods are in excellent agreement, Fig.~\ref{fig:results}{d}.

The transverse field Ising model at $T=0$ has a quantum critical point at $h_x=|c|$ \cite{Sachdev_2011}. The system is ferromagnetically ordered ($m_z\ne0$) for $|c|>h_x$, whereas it is quantum disordered for $|c|<h_x$ (hence  $m_z=0$). This provides an ideal test case. 
The magnetization calculated via $\overline{G_2}$ agrees well with the exact solution \cite{Ising_spin_spin_correlators,Pfeuty_ising}, Fig.~\ref{fig:results}{e}. However, there is some disagreement very close to the quantum critical point. The agreement between the magnetization calculated from $\overline{G_2}$ and directly from DMRG shows that the disagreement with the exact result is due to the limitations of DMRG near the critical point rather than any issue with the generalised Maxwell relation. Thus, we conclude that the generalised Maxwell relation holds in the most stringent of conditions, close to a critical point.

Interestingly, close to the quantum critical point the $\partial\overline{G_2}/\partial h_z$ displays a `lambda' anomaly, Fig.~\ref{fig:results}{e} (inset), providing a clear signature of the continuous phase transition and allowing the measurement of critical exponents (see Supplementary Information). This is because the integrated correlation function is a thermodynamic function and therefore provides direct thermodynamic information. As $\partial\overline{G_2}/\partial h_z=(\partial^2 F/\partial c\,\partial h_z)_{T,\mathcal{V},N}$ is a second derivative of the free energy, it should be expected to display a lambda anomaly at the critical point---exactly as we find in our calculations. Thus, measurements of integrated correlation functions provide a new method to characterise phase transitions in quantum many-body systems.

This provides a useful approach for probing the physics of engineered quantum systems, such as 2D materials. Conventional thermodynamics is challenging is such systems---for example the heat capacity is typically dominated by the substrate. Relevant correlation functions might be measured via local electron tunnelling spectroscopic probes \cite{Leeuwenhoek_2020,Niu_1995,Byers_1995}.

\textbf{\emph{Heat capacity}} ($Y\!=\!-C_{\mathcal{V}}/T$, $X\!=\!X'\!=\!T$).
Like many of the most frequently measured thermodynamic quantities, the heat capacity is a second derivative of the free energy. {In this case, the generalized Maxwell relation \eqref{eq:2Maxwell-generic} between the (constant volume) heat capacity, $C_{\mathcal{V}}=-T\left( \partial^2 F/ \partial T^2\right)_{N,\mathcal{V},c}$, and the integrated correlation function, $\overline{G_2}$, is given by Eq.~\eqref{eq:heat_capacity_table}, Table \ref{tab:Maxwell_all}.}

To test this Maxwell relation we computed $C_{\mathcal{V}}$ for the transverse field Ising model from: (i) the calculated $\overline{G_2}$ via Eqs.~\eqref{Maxwell-integral} and \eqref{eq:heat_capacity_table} with $c_0=0$; and (i) the variance of the energy \cite{Callen_book}.
Both calculations are in excellent agreement, Fig.~\ref{fig:results}{f}. Similar agreement with standard thermodynamics is found for the isothermal compressibility of the Bose gas (Supplementary Information).
Both results confirm the accuracy of the generalised Maxwell relation approach for second derivatives of the free energy.

~

\noindent\textbf{Conclusions}\\
~
The generalised Maxwell relations, derived above, provide a powerful, universal approach to measuring the thermodynamics of quantum simulators and other systems where it is significantly easier to measure correlation functions than to perform traditional thermodynamic measurements. 
These relations bridge the gap between microscopic measurements and macroscopic properties by linking observables such as pressure, entropy, and heat capacity to derivatives of correlation functions, thereby extending the traditional paradigm of macroscopic thermodynamics.

While, in this paper, we have limited our discussion to pairwise interactions and pair correlation functions, Maxwell relations can be straightforwardly generalised to higher-body interactions with the pair correlation function replaced by the appropriate higher order correlation function.

Many quantum simulators allow the interaction strength, $c$, to be continuously varied, enabling the integrals required to evaluate thermodynamic properties (equations~\eqref{Maxwell-integral} and \eqref{eq:2Maxwell-integral}) to be evaluated from experimental data. In conventional materials this is typically not possible. Nevertheless, as shown above, the integrated correlation function, $\overline{G_2}$, is itself a thermodynamic variable. Thus, direct measurement of local correlation functions will enable thermodynamic measurements in systems, such as 1D and 2D materials, where traditional thermodynamic measurements are not possible, for example, due to such measurements being dominated by a substrate. 

Conversely, in many materials thermodynamic measurements are straightforward while the most interesting correlation functions cannot be measured. For example, magnetic systems that cannot be grown in large enough single crystals to enable inelastic neutron scattering. The generalised Maxwell relations could provide significant insights into correlations in such materials by 
inverting our approach, i.e., by extracting integrated correlation functions from traditional measurements of thermodynamic quantities.

Finally, we note that thermometry is difficult in some quantum simulators, e.g., quantum gas microscopes. The generalised Maxwell relations {may} offer a solution to this by determining the temperature from measurements of the integrated correlation functions.

%%%%%%%%%%%%%%%%%%%%%%%%%%%%%%%%%%%%%%%%%%%%%%%%%%
%\bibliography{Bibliography.bib}


%%%%%%%%%%%%%%%%%%%%%%%%%%%%%%%%%%%%%%%%%%%%%%%%%%%

~
 
\noindent\textbf{\large{Methods}}\\

\noindent\textbf{Maxwell relations}\\
\noindent 
The generalised Maxwell relations involving correlation functions are derived here in the canonical formalism of statistical mechanics. The Helmholtz free energy, $F$, is  given by
\begin{equation}
F= -k_BT \ln \Big(\text{Tr}e^{-\hat{H}/k_BT} \Big).
\end{equation}
Applying the Hellmann-Feynman theorem \cite{hellmann1933rolle,Feynman:1939,Kheruntsyan2003} to the general Hamiltonian for pairwise interactions, equation~\eqref{eq:Hamiltonian-general}, one finds that
\begin{align}
  \frac{\partial F}{\partial c} &= \frac{1}{Z} \text{Tr}\Big(e^{-\hat{H}/k_BT} \frac{\partial \hat{H}}{\partial c}\Big)=\overline{G_2} ,
\label{dF_dchi_Methods}
\end{align}
where $Z = \text{Tr}\exp(-\hat{H}/k_BT)$ is the partition function.
Whence, the generic form of Maxwell relations, equation~\eqref{Maxwell-generic}, follows from the commutativity of mixed second derivatives, equation~\eqref{eq:mixed}.

In the Supplementary Information we show how the same Maxwell relations can be derived (i) without resorting to the Hellmann-Feynman theorem but instead starting from the observation that $c$ and $\overline{G_2}$ are conjugate thermodynamic parameters; and (ii) in other thermodynamic ensembles.
 \\

\noindent\textbf{Lieb-Liniger model}\\
The Lieb-Liniger model of a uniform gas of $N$ bosons interacting in 1D via pairwise contact ($\delta$-function) interaction is specified by equation~\eqref{eq:Hamiltonian-general}, with 
\begin{align}
	&\hat{H}_0	=  -\frac{\hbar^{2}}{2m}  \int_{0}^{\mathcal{V}} dx\,  \hat{\Psi}^{\dagger}(x) \frac{\partial^2 \hat{\Psi}(x)}{\partial x^2}, \\
    &\hat{\overline{G_2}}= \int_{0}^{\mathcal{V}}  dx\, \hat{\Psi}^{\dagger}(x) \hat{\Psi}^{\dagger}(x) \hat{\Psi}(x) \hat{\Psi}(x).\label{eq:H-LL}
\end{align}

In the strongly interacting limit the integrated correlation function $\overline{G_2}(c)$ is calculated  at $T=0$ using DMRG following \cite{Simmons2020}.The integral form of the generalised Maxwell relation {for pressure, Eq.~\eqref{eq:pressure_table},} then allows us to compute $P(c)$ {(shown in Fig.~\ref{fig:results}a)} after  setting $c_0\to\infty$, which corresponds to the strongly interacting Tonks-Girardeau limit of the 1D Bose gas, for which the pressure is identical to that of the 1D non-interacting {spinless} Fermi gas (see, e.g., \cite{kerr2024analytic}).

~

\noindent\textbf{Yang-Gaudin model}\\
\noindent The Yang-Gaudin model \cite{Yang1967,gaudin1967systeme,Guan2013} is similar to the Lieb-Liniger model, except that it describes a system of interacting spin-1/2 fermions in 1D. The corresponding Hamiltonian can be written as a sum of two terms:
\begin{align}
	&\hat{H}	_0= -\sum_{\sigma=\uparrow,\downarrow}\frac{\hbar^{2}}{2m}  \int_{0}^{\mathcal{V}}  dx\,  \hat{\Psi}_{\sigma}^{\dagger}(x) \frac{\partial^2 \hat{\Psi}_{\sigma}(x)}{\partial x^2}, \\ 
	&\hat{\overline{G_2}}=  \int_{0}^{\mathcal{V}}  dx\, \hat{\Psi}_{\uparrow}^{\dagger}(x) \hat{\Psi}_{\downarrow}^{\dagger}(x) \hat{\Psi}_{\downarrow}(x) \hat{\Psi}_{\uparrow}(x),\label{eq:H-YG}
\end{align}
where the field operators $\hat{\Psi}^{(\dagger)}_{\sigma}(x)$ annihilate (create) a fermion with spin $\sigma=\uparrow,\downarrow$ at position $x$.\\

\noindent\textbf{Fermi- and Bose-Hubbard models}\\
\noindent The  Fermi-Hubbard model is specified by
\begin{align}
	&\hat{H}_0= -t \sum_{j=1}^{\mathcal{V}} \sum_{\sigma=\uparrow,\downarrow} \!\left( \hat{c}^\dagger_{j,\sigma} \hat{c}^{}_{j+1,\sigma} + \mathrm{h.c.}\right),\\
	&\hat{\overline{G_2}}= \sum_{j=1}^{\mathcal{V}} \hat{c}^{\dagger}_{j,\uparrow} \hat{c}_{j,\uparrow} \hat{c}^{\dagger}_{j,\downarrow} \hat{c}_{j,\downarrow}.
\end{align}
All results presented for the Fermi-Hubbard model are calculated via DMRG and scaled to the thermodynamic limit. We keep a truncated bond dimension of up to $m=1000$, and scale our results to the $m\rightarrow\infty$ limit using the variance in the internal energy.

The Bose-Hubbard model is given by
\begin{align}
	&\hat{H}_0= \!- t \sum_{j=1}^{\mathcal{V}} \!\left( \hat{b}^\dagger_j \hat{b}^{}_{j+1} + \mathrm{h.c.}\right),\\ 
	&\hat{\overline{G_2}}= \sum_{j=1}^{\mathcal{V}}   \hat{b}^{\dagger}_j \hat{b}_j \hat{b}^{\dagger}_j \hat{b}_j
	,
\end{align}
where $\hat{b}^{(\dagger)}_j$ annihilates (creates) a boson on site $j$. All results for the Bose-Hubbard model are calculated via DMRG for 100 lattice sites with a bond dimension of $m=100$. The bond dimension, which controls the accuracy of the numerical results, can be much lower than in the Fermi-Hubbard model due to the lower degree of entanglement in the Bose-Hubbard model.

~

\noindent\textbf{Transverse field Ising model}\\
The  transverse field Ising model is given by
\begin{align}
	& \hat{H}_0=h_x\sum_{j=1}^{\mathcal{V}}\hat{S}_j^x , \label{Ising-field}\\
	&\hat{\overline{G_2}}= \sum_{j=1}^{\mathcal{V}-1}\hat{S}_j^z\hat{S}_{j+1}^z.
	\label{Ising-interaction}
\end{align}

We calculate {the finite-temperature thermal equilibrium state} of the transverse Ising model using the ancilla approach to DMRG  \cite{ancilla_DMRG}. We start from an infinite temperature state which is maximally entangled between the physical sites and ancilla states, and evolve in imaginary time using time evolving block decimation and a second order Trotter gate decomposition \cite{ancilla_DMRG}.
To avoid numerical errors in the heat capacity (Fig. \ref{fig:results}{f}), arising from calculating higher order numerical derivatives at small time steps, we employ a Savitzky-Golay filter to smooth $\overline{G_2}$ at low temperatures \cite{Savitzky_goly_filter}.

The derivative of $(\partial \overline{G_2}/\partial h_z)$, shown in the inset of Fig.~\ref{fig:results}{e}, cannot be directly calculated for the transverse Ising model due to the $\mathcal{Z}_2$ symmetry ($S^z_j\rightarrow -S^z_j$) of  $\overline{G_2}$. Thus,  we explicitly break this symmetry  by introducing a longitudinal field, i.e.,  adding a term $h_z\sum_{j=1}^{\mathcal{V}}\hat{S}_j^{z}$ to  Hamiltonian \eqref{Ising-field},
and numerically evaluate $\partial \overline{G_2}/\partial h_z=\lim_{h_z\rightarrow 0}[\partial \overline{G_2}(h_z)/\partial h_z]$.

~

\noindent\textbf{\large{Data availability}}\\
All data presented here will be made available upon publication.
%is available from \textcolor{red}{https://... ADD LINK}

~

\noindent\textbf{\large{Acknowledgments}}\\~
We acknowledge the stimulating and insightful discussions with I. Bloch, P. Jacobson and L. Pollet. This work was supported through Australian Research Council Discovery Project Grant Nos. DP190101515, DP230100139 and DP240101033.

~

\noindent\textbf{\large{Author contributions}}\\~
All authors contributed to the development of the method.
F.R. and R.S.W. carried out and analysed the calculations for fermionic/spin and bosonic models, respectively. 
H.L.N., B.J.P. and K.V.K. directed the research.
B.J.P. and K.V.K. conceived and planned the research.

~

\noindent\textbf{\large{Competing interests}}\\~
The authors declare no competing interests.

~

\noindent\textbf{\large{Additional information}}\\~
\textbf{Supplementary Information} Supplementary Information is available for this paper.

~

\textbf{Correspondence and requests for materials} should be addressed to
B. J. Powell or K. V. Kheruntsyan.

\clearpage

\section{Supplementary Information}

\renewcommand\thefigure{S\arabic{figure}}
\setcounter{figure}{0}  
\renewcommand\thetable{S\arabic{table}}
\setcounter{table}{0}  
\renewcommand\theequation{S\arabic{equation}}
\setcounter{equation}{0}  
\renewcommand\thesection{S\arabic{section}}

\subsection{S1. Thermodynamic potentials and identities for Maxwell relations involving correlations}

In this section we derive generalised Maxwell relations for systems with arbitrary two-body interactions  without resorting to the Hellmann-Feynman theorem. We first note that the integrated correlation function, $\overline{G_2} $, is an \emph{extensive} thermodynamic parameter that characterizes the variation of the internal energy of the system, $U=\langle \hat{H} \rangle$, with the interaction strength $c$, which is an \emph{intensive} parameter conjugate to $\overline{G_2}$. The extensivity of $\overline{G_2}$ follows immediately from the fact that $c\, \overline{G_2}$ is simply the expectation value of the interaction part of the Hamiltonian, $\langle \hat{H}_{\mathrm{int}}\rangle=c\, \langle \hat{\overline{G_2}} \rangle$, which itself must be extensive because the overall Hamiltonian energy and the kinetic energy part alone are also extensive. For example, for a uniform 1D Lieb-Liniger gas  this can also be seen explicitly as $\overline{G_2}=\mathcal{V} \langle \hat{\Psi}^{\dagger}\hat{\Psi}^{\dagger}\hat{\Psi}\hat{\Psi}\rangle=\mathcal{V}n^2g^{(2)}$, where $\mathcal{V}$ is extensive, whereas $n$ and $g^{(2)}$ are intensive.

Next, we make the dependence of the free energy on $c$ explicit, $F=F(T,\mathcal{V},N,c,\dots)$. Accordingly, the variation of the generalised Helmholtz free energy can be written as  
\begin{equation}
dF=-SdT-Pd\mathcal{V}+\mu dN+\overline{G_2}dc+\dots,
\label{eq:dF}
\end{equation}
where $\mu =(\partial F/\partial N)_{T,\mathcal{V},c}$ is the chemical potential, and $\overline{G_2}=(\partial F/\partial c)_{T,\mathcal{V},N}$ which we note agrees with equation~\eqref{dF_dchi_Methods} except that now it is derived without relying on the Hellmann-Feynman theorem. From equation~\eqref{eq:dF}, one can derive all Maxwell relations as usual.

\subsection{{S2. Explicit integral forms of generalised Maxwell relations}}

{In  table \ref{tab:Maxwell_int}, we summarise the explicit integral forms, Eqs.~\eqref{Maxwell-integral} and \eqref{eq:2Maxwell-integral}, of generalized Maxwell relations presented in Table \ref{tab:Maxwell_all}.}
%\newcounter{tblEqCounter} %create a counter
%\setcounter{tblEqCounter}{\theequation} %at the start of the table, set the counter to equation numbering
\begin{table}[h!]
\caption{Integral forms of the generalised Maxwell relations involving the pair correlation $\overline{G_2}$.}
\label{tab:Maxwell_int}
\centering
%\scriptsize
\begin{tabular}{llc}
\toprule
Thermodynamic \\quantity & Integral form of the Maxwell relation for $\overline{G_2}$ \\
\midrule
Pressure & ${\displaystyle {\displaystyle P(c)=P(c_0) -\int_{c_0}^{c} \left(\frac{\partial \overline{G_2}(c')}{\partial \mathcal{V}}\right)_{c',T,N}dc'.}}$ & %\{eq:pressure_table}
\\ %set the equation number 
Entropy & ${\displaystyle {\displaystyle S(c)=S(c_0)
	-\int_{c_0}^{c} \left(\frac{\partial \overline{G_2}(c')}{\partial T}\right)_{c',\mathcal{V},N}dc'}}$ & %\numberTblEq{eq:entropy_table}
\\
$\begin{array}{c}
\vspace{-0.2cm}\\
\hspace{-0.2cm}\text{\,\,Chemical}\\
\hspace{-0.1cm}\text{potential}\\
\vspace{-0.3cm}
\end{array}$ & ${\displaystyle {\displaystyle \mu(c)=\mu(c_0) +\int_{c_0}^{c} \left(\frac{\partial \overline{G_2}(c')}{\partial N}\right)_{c',T,\mathcal{V}}dc'}}$ & %\numberTblEq{eq:chemical_potential_table}
\\ 
Magnetization & ${\displaystyle {\displaystyle \bm{m}(c)=\bm{m}(c_0)-\int_{c_0}^{c} \bm{\nabla_{\bm{h}}}\overline{G_2}(c')dc'}}$ & %\numberTblEq{eq:magnetization_table}
\\ 
Heat capacity & ${\displaystyle {\displaystyle C_{\mathcal{V}}(c) \!=\! C_{\mathcal{V}}(c_0) \!-\!T \!\int_{c_0}^{c}\! \left(\frac{\partial^2 \overline{G_2}(c')}{\partial T^2} \right)_{\!\!c',\mathcal{V},N} \!dc'}}$ & %\numberTblEq{eq:heat_capacity_table}
\\ 
$\begin{array}{c}
\vspace{-0.2cm}\\
\hspace{-0.8cm}\text{\,\,Isothermal}\\
\hspace{-0.1cm}\text{compressibility}\\
\vspace{-0.3cm}
\end{array}$
 & ${\displaystyle {\displaystyle 
\kappa_{T}^{-1}(c) \!=\! \kappa_{T}^{-1}(c_0) \!+\!\mathcal{V} \!\int_{c_0}^{c} \!\left(\frac{\partial^2 \overline{G_2}(c')}{\partial T^2} \right)_{\!\!c',T,N} \!dc'}}$ & %\numberTblEq{eq:compressibility_table}
 \\ \bottomrule
\end{tabular}
\end{table}

\subsection{S3. Other thermodynamic ensembles}

According to the standard formalism of thermodynamics (see, e.g., \cite{Callen_book,Alberty2002}), the fundamental equation~\eqref{eq:dF} itself can be obtained from the Euler equation for the generalised internal energy of the system,
\begin{equation}
U(S,\mathcal{V},N,\overline{G_2},\dots)=TS-P\mathcal{V}+\mu N-c \,\overline{G_2}+\dots,
\end{equation}
which is a function of only extensive parameters, by means of the following Legendre transform:
\begin{equation}
F=U-TS+c\,\overline{G_2}.
\label{eq:F_Legendre}
\end{equation}

The differential of $U$ is given by  
\begin{equation}
dU=TdS-Pd\mathcal{V}+\mu dN-c\, d\overline{G_2},
\label{eq:dU}
\end{equation}
where $T\!=\!(\partial U/\partial S)_{\mathcal{V},N,\overline{G_2}}$, $P\!=\!-(\partial U /\partial \mathcal{V})_{S,N,\overline{G_2}}$, $\mu\!=\!(\partial U/\partial N)_{S,\mathcal{V},\overline{G_2}}$, and $c \!=\! -\left(\partial U/\partial\overline{G_2}\right)_{S,\mathcal{V},N}=T\left(\partial S/\partial\overline{G_2}\right)_{U,\mathcal{V},N}$. 
 Equations~\eqref{eq:F_Legendre} and \eqref{eq:dU} lead directly to \eqref{eq:dF}.

An equation similar to Eq.~\eqref{eq:dF} can be written  for the grand-canonical thermodynamic potential $\Omega=F-\mu N$, 
\begin{equation}
d\Omega=-SdT-Pd\mathcal{V}-Nd\mu+\overline{G_2}dc+\dots,
\label{eq:dO}
\end{equation}
with $\Omega=\Omega(T,\mathcal{V},\mu,c,\dots)$. If there are no additional relevant parameters (represented by the ellipsis) then for homogeneous systems $\Omega=-P\mathcal{V}$ and equation~\eqref{eq:dO} can be further simplified to 
\begin{equation}
\mathcal{V}dP-SdT-Nd\mu+\overline{G_2}dc=0,
\label{eq:Gibb-Duhem}
\end{equation}
which takes the role of the generalised Gibbs-Duhem relation. The generalised Gibbs-Duhem relation implies that among the four intensive parameters $\{P,T,\mu,c\}$ only three are independent, and hence the dependence of the fourth parameter on the other three takes the role of the thermodynamic equation of state, such as $P=P(T,\mu,c)$. For explicit examples of such equations of state for the uniform 1D Bose gas, see a recent review in Ref.~\cite{kerr2024analytic}.

\begin{figure}[tbp] 
    \includegraphics[width=8.0cm]{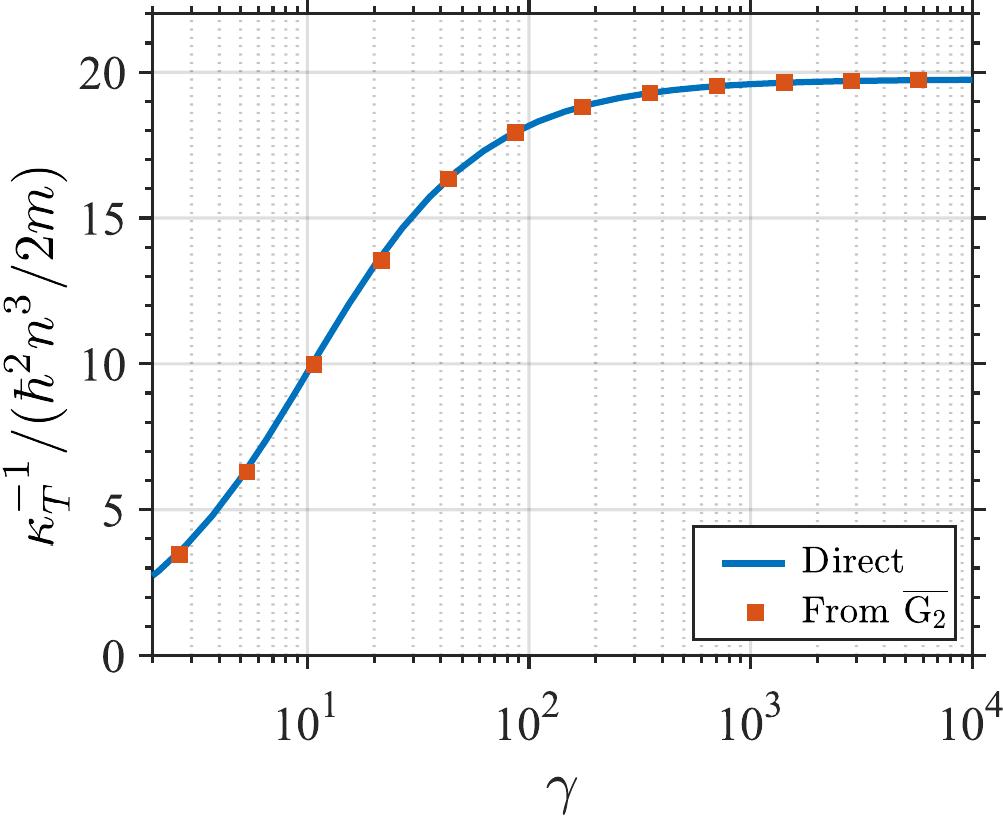}
    \caption{Inverse isothermal compressibility ($\kappa_T^{-1}$) of the Lieb-Liniger model at $T=0$ K in the strongly interacting regime ($\gamma \!>\! 1$). Calculations obtained directly and from numerical $\overline{G_2}$ are achieved using the Bethe ansatz state solutions for the ground state \cite{Lieb-Liniger-I}, which are exact. 
    }
    \label{fig:LiebLiniger_InverseCompressibility}
\end{figure}

\subsection{S4. Inverse of isothermal compressibility for the Lieb-Liniger model ($Y=\kappa_T^{-1}/\mathcal{V}$, $X=X'=\mathcal{V}$)}

The inverse of isothermal compressibility, $\kappa_T=-\mathcal{V}^{-1}\left( \partial \mathcal{V}/\partial P\right)_{N,T,c} $, can be expressed through the Helmoltz free energy via $\kappa_T^{-1}=\mathcal{V} \left( \partial^2 F/\partial \mathcal{V}^2\right)_{T,N,c}$, and therefore the generic equation~\eqref{eq:2Maxwell-generic} {leads to the Maxwell relation given as Eq.~\eqref{eq:compressibility_table} in Table \ref{tab:Maxwell_all}.}

We  evaluate $\kappa_T^{-1}$, {using Eqs.~\eqref{eq:2Maxwell-integral} and \eqref{eq:compressibility_table},} for the strongly interacting ($\gamma \! > \! 1$) Lieb-Liniger model at $T=0$,   where $\overline{G_2}$ is calculated using the Hellmann-Feynman theorem \cite{Kheruntsyan2003}.
This is in excellent agreement with direct calculation of $\kappa_T^{-1}$ from the ground state energy using the Bethe ansatz \cite{Lieb-Liniger-I}, Fig.~\ref{fig:LiebLiniger_InverseCompressibility}.

\subsection{S5. Entropy for the Yang-Gaudin model ($Y \equiv -S$, $X\equiv T$)}

The entropy in the canonical ensemble is defined via $S=- \left( \partial F/\partial T\right)_{\mathcal{V},N,c}$ and therefore the generalised Maxwell relation between $S$ and $\overline{G_2}$ {acquires the form of Eq.~\eqref{eq:entropy_table}, Table \ref{tab:Maxwell_all}.}

To illustrate the utility of this Maxwell relation in the Yang-Gaudin model \cite{Yang1967,gaudin1967systeme,Guan2013}, we utilise the known  
approximate analytic expressions for the normalised local pair correlation function \cite{patu2016thermodynamics},
\begin{align}
g^{(2)}_{\uparrow\downarrow}(0) &= 4 \frac{\langle \hat{\Psi}_{\uparrow}^{\dagger}(x) \hat{\Psi}_{\downarrow}^{\dagger}(x) \hat{\Psi}_{\downarrow}(x) \hat{\Psi}_{\uparrow}(x) \rangle }{n^2 } 
\\
&=(1-\mathcal{P}^2)\frac{\langle \hat{\Psi}_{\uparrow}^{\dagger}(x) \hat{\Psi}_{\downarrow}^{\dagger}(x) \hat{\Psi}_{\downarrow}(x) \hat{\Psi}_{\uparrow}(x) \rangle }{n_{\uparrow} n_{\downarrow }} ,
\label{Yang-Gaudin-correlation-def}
\end{align}
where $n=n_{\uparrow}+n_{\downarrow}$ is the total uniform density, $n_{\uparrow(\downarrow)}=\langle \hat{\Psi}^{\dagger}_{\uparrow(\downarrow)}(x)\hat{\Psi}_{\uparrow(\downarrow)}(x) \rangle $ is the density in the spin up (down) component, and $\mathcal{P}=(n_{\uparrow} - n_{\downarrow})/n$ is the polarisation.  The local pair correlation $g^{(2)}_{\uparrow\downarrow}(0)$ can be found through a relation with Tan's contact parameter $\mathcal{C}$, namely
$g^{(2)}_{\uparrow\downarrow}(0) = (4/c^2 n^2)\,\mathcal{C}$ \cite{Tan_1,Tan_2,Zwerger_Tan_2011} and hence $\overline{G_2} = (4\mathcal{V}/c^2 n^2)\,\mathcal{C}$, and the known result for $\mathcal{C}$ itself
\cite{patu2016thermodynamics,he2016universal}.

At sufficiently high temperatures, $k_B T \gg \hbar^2 c^2/2m$, these results allow us to express $g^{(2)}_{\uparrow\downarrow}(0)$ as 
\begin{equation}
\begin{split}
\begin{aligned}
    g^{(2)}_{\uparrow\downarrow}(0) = (1 - \mathcal{P}^2)  \bigg\{ 1 \!-\! \frac{\sqrt{\pi}  \gamma}{\sqrt{2 \tau}} e^{\frac{ \gamma^2}{2\tau}} \left[ 1 - \mathrm{erf}\left( \frac{\gamma}{\sqrt{2\tau}}\right) \right]\bigg\},
\end{aligned}
\end{split}
\label{g2-Yang-Gaudin-at-high-T}
\end{equation}
where we have introduced a dimensionless interaction strength $\gamma\!=\!2cm/\hbar^2n$ for constant total density $n$, and a dimensionless temperature $\tau\!=\!2 m k_B T / \hbar^2 n^2$.
Simplifying this expression further using the lowest order Taylor expansion (such that the resulting entropy may be compared with the similar analytic result derived in Ref.~\cite{Watson2024}  for the Lieb-Liniger model), {we then use Eqs.~\eqref{Maxwell-integral} and \eqref{eq:entropy_table}} to obtain: 
\begin{equation}\label{eq:YangGaudin_S}
	S \simeq S_{\mathrm{IFG}} -  \frac{N (1-\mathcal{P}^2)}{16\hbar } \sqrt\frac{\pi m}{2k_B T^3} nc^2.
\end{equation}
Here, $S_{\mathrm{IFG}}\equiv S(c=0)$ is the entropy of an ideal Fermi gas (see, e.g., Ref. \cite{kerr2024analytic}).
To the best of our knowledge, this approximate analytic result for the entropy of the Yang-Gaudin model has not been derived before.

~

\subsection{S6. Exact calculation of the magnetization of the transverse field Ising model at $T=0$}

The spin-spin correlation functions for the transverse field Ising model have been extensively studied, with the nearest neighbour longitudinal spin-spin correlation given by \cite{Pfeuty_ising}
\begin{equation}
\label{eq:ising_g2}\overline{G_2} = \frac{1}{4\pi}\int_0^\pi dk \ \Lambda_k^{-1}\cos(k)+ \frac{c}{8h_x\pi}\int_0^\pi dk \ \Lambda_k^{-1}  ,\end{equation}
where $\Lambda_k$ is the energy of the non-interacting fermions after a Jordan-Wigner transformation is performed, given by
\begin{equation}\Lambda_k= \sqrt{1+\bigg{(}\frac{c}{h_x}\bigg{)}^2+\frac{c\cos(k)}{h_x}} .\end{equation}
The derivative of equation~\eqref{eq:ising_g2} with respect to $h_x$ is given by 
\begin{align}\frac{\partial\overline{G_2}}{\partial h_x} &= \frac{1}{4\pi}\int_0^\pi dk \ \frac{\partial \Lambda_k^{-1}}{\partial h_x}\bigg{(}\cos(k)+\frac{c}{h_x}\bigg{)}\nonumber
\\&+ \frac{c}{8h_x^2\pi}\int_0^\pi dk \ \Lambda_k^{-1}\label{eq:g2_magx_derivative}.  
\end{align}
Combining Eq.~\eqref{eq:g2_magx_derivative} with {Eqs.~\eqref{Maxwell-integral} and \eqref{eq:magnetization_table} of the main text} allows the transverse magnetization at zero temperature to be calculated, Fig.~\ref{fig:results}{d}.

\subsection{S7. Scaling of $\partial \overline{G_2}/\partial h_x$ near the quantum critical point of the transverse field Ising model}

Evaluating $\partial \overline{G_2}/\partial h_z$ to measure the longitudinal magnetisation, $m^z$, for the Ising model, results in a `lambda' anomaly as shown in the inset of Fig. \ref{fig:results}e. As the longitudinal magnetization serves as the order parameter in this model,  the order parameter critical exponent, $\beta$, can be extracted from $\partial \overline{G_2}/\partial h_z$ close to the critical point.

\begin{figure}[tbp] 
    \includegraphics[width=8.0cm]{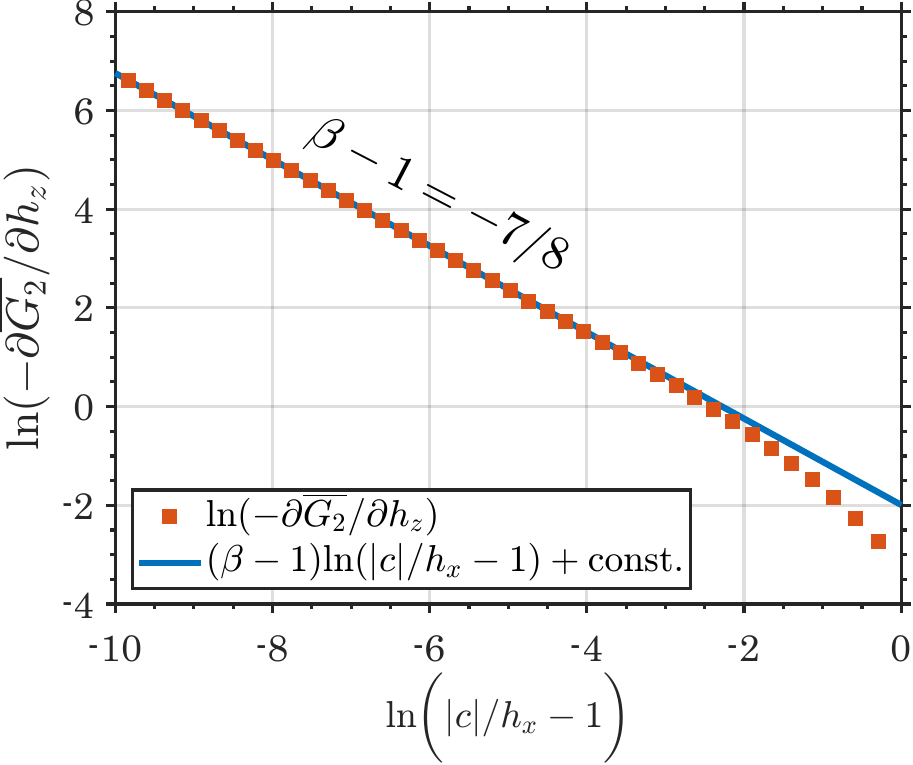}
    \caption{\label{fig:critical_exponent}  Zero temperature $\partial \overline{G_2} / \partial h_x$ displays scaling behaviour near the critical point of the transverse field Ising model.  Linear regression reveals the scaling law behaviour follows $-\partial G_2 / \partial h_x \propto [ |c|/h_x-1 ]^{\beta-1}$ where $\beta= 1/8$ is the order parameter critical exponent. $\partial \overline{G_2} / \partial h_x$ is calculated exactly using Eq. \eqref{eq:critical_exponent_equation}. }
\end{figure}

The longitudinal magnetization for the transverse Ising model is analytically given by $m_z=(1-(|c|/h_x)^{-2})^{\beta}$ \cite{Pfeuty_ising}. Using the Maxwell relation given by Eq. \eqref{eq:magnetization_table}, $-\partial \overline{G_2}/\partial h_z=\partial m_z/\partial c$, it can be shown that $\mathrm{ln}(-\partial \overline{G_2}/\partial h_z)$ is exactly given by 
\begin{align}\mathrm{ln}\bigg{(}-\frac{\partial \overline{G_2}}{\partial h_x}\bigg{)}=&-3\mathrm{ln}(|c|/h_x)-\mathrm{ln}(4)\nonumber
\\&+(\beta-1)\ln\bigg{(}1-(|c|/h_x)^{-2}\bigg{)}.\label{eq:critical_exponent_equation}
\end{align} 
Near the critical point ($|c|=h_x$) we have $\lim_{|c|/h_x\rightarrow 1}\mathrm{ln}(|c|/h_x)=0$ and $1-(|c|/h_x)^{-2}\simeq 2(|c|/h_x-1)$; hence the critical exponent can then be extracted from a log-log plot of $\partial \overline{G_2}/\partial h_z$. This is illustrated in Fig. \ref{fig:critical_exponent}, where, as the critical point is approached, $\partial \overline{G_2}/\partial h_z$ displays linear scaling behaviour and the critical exponent can be extracted. We find excellent agreement with the textbook result  $\beta=1/8$ \cite{Pfeuty_ising,Sachdev_2011}.

\end{document}